\documentclass[
  aps,
  prd,
  preprint,
//  linenumbers,
  floatfix,
  nofootinbib,
  eqsecnum,
  superscriptaddress
]{revtex4-2}

\usepackage[T1]{fontenc}		
\usepackage{amsmath,amsfonts,amssymb,amstext,mathrsfs}
\usepackage{mathpazo}
\usepackage{upgreek}

\usepackage[todonotes={textsize=scriptsize, backgroundcolor=red!20}]{changes}
\definechangesauthor[name=Rafał, color=orange!70!black]{RS}
\definechangesauthor[name=MKG, color=purple!70!black]{MKG}
\usepackage{graphicx}
\usepackage{placeins}
\usepackage{booktabs}

\makeatletter
\long\def\@makecaption#1#2{%
  \par
  \vskip\abovecaptionskip
  \begingroup
    \small\rmfamily
    \samepage
    \flushing
    \let\footnote\@footnotemark@gobble
    \@make@capt@title{#1}{#2}\par
  \endgroup
  \vskip\belowcaptionskip
}
\makeatother

\begin{document}

\title{Sensitivity of the photon-induced processes to the proton charge radius}


\author{Nikhil Krishna}
 \email{Nikhil.Krishna@ifj.edu.pl}
\affiliation{Institute of Nuclear Physics Polish Academy of Sciences, Radzikowskiego 152, PL-31-342 Krak\'ow, Poland}

\author{Mariola K{\l}usek-Gawenda}
 \email{Mariola.Klusek@ifj.edu.pl}
\affiliation{Institute of Nuclear Physics Polish Academy of Sciences, Radzikowskiego 152, PL-31-342 Krak\'ow, Poland}

\author{Rafa{\l} Staszewski} 
\email{Rafal.Staszewski@ifj.edu.pl}
\affiliation{Institute of Nuclear Physics Polish Academy of Sciences, Radzikowskiego 152, PL-31-342 Krak\'ow, Poland}

\date{\today}

\begin{abstract}

We study the exclusive production of dileptons in proton--proton collisions as a probe of the proton charge-radius scale. Using a dipole form factor model, we compare the conventional choice of $\Lambda^2=0.71$~GeV$^2$ with PDG test scenarios corresponding to $r_p=0.8751 \pm 0.0061$~fm and $r_p=0.84087 \pm 0.00039$~fm, representative of the historical large-radius from CODATA and smaller-radius from muonic-hydrogen determinations, respectively. The sensitivity is greatest at large dilepton invariant masses and forward/backward rapidity. Fitting to the current ATLAS and CMS data within the adopted model gives $\Lambda^2 = 0.465 \pm 0.056~\mathrm{GeV}^2$, corresponding to an effective
radius $r_p = 1.002 \pm 0.038~\mathrm{fm}$, which indicates non-trivial sensitivity on the proton radius scale, but is not yet a definitive solution to the proton radius puzzle. While this result depends on the theoretical model used in the calculation,  it shows that exclusive dilepton production at LHC can provide complementary model-dependent information on the proton electromagnetic forms factor.

\end{abstract}

\maketitle

\section{Introduction}
\label{sec:intro}
The proton charge radius is a fundamental quantity in hadronic physics. Historically, its determination via electron scattering and atomic spectroscopy has served as a precision benchmark. However, the significantly smaller value extracted from muonic hydrogen spectroscopy gave rise to the well-known proton radius puzzle. It was a motivation to renew both experimental methodologies and theoretical interpretations \cite{RevModPhys.94.015002,Carlson:2015jba,pohl2010size,Antognini:2013txn}.

In parallel, high-energy proton--proton collisions provide a complementary avenue through which the electromagnetic structure of the proton may be probed. In particular, exclusive photon-induced dilepton production, $pp \to pp \ell^+ \ell^-$, proceeds via the fusion of two quasi-real photons emitted by the incoming protons. Such processes are naturally described within the Equivalent Photon Approximation (EPA), which has constituted a standard theoretical framework for two-photon physics for several decades \cite{Budnev:1975poe}. In the context of proton-proton collisions, quantitatively reliable predictions require adequate consideration of the influence of the finite size of the emitting proton, especially in the case of the precise comparison with the LHC data \cite{Dyndal:2014yea}.

The electromagnetic structure of the proton is encoded in the Sachs electric and magnetic form factor, $G_E(Q^2)$ and $G_M(Q^2)$, which represent the standard quantities employed in the analysis of elastic electron-proton scattering~\cite{Sachs:1962zzc,A1:2010nsl,Perdrisat:2006hj}. Recent high-precision measurements and exclusive analyses have substantially advanced our knowledge of these form factors, particularly in the low-$Q^2$ region, which is crucial to charge radius extraction \cite{A1:2010nsl,Perdrisat:2006hj}. Within phenomenological applications, it is common to adopt a dipole parametrization and to relate its parameter $\Lambda^2$ to an effective proton charge radius \cite{RevModPhys.94.015002,A1:2010nsl,Perdrisat:2006hj}. This approach provides a convenient theoretical framework for examining how variations in the proton radius manifest in observables measured in hadron collisions.

In the present work, we examine the sensitivity of exclusive dilepton production at the LHC to the proton-radius parameter within the aforementioned dipole framework. The analysis is carried out within two complementary theoretical approaches: the momentum-space treatment of the $2 \to 4$ process and the impact parameter representation. The latter permits the inclusion of absorptive corrections via a survival factor. Although the primary physics goal of this work concerns the dependence of the cross section observables on the proton radius, a quantitatively reliable analysis requires the preservation of the full electromagnetic structure of the proton in the photon flux. Accordingly, all final results are obtained employing the complete Sachs form factor treatment, while sensitivity to the proton radius is investigated through systematic variation of the dipole parameter. 

It is demonstrated that sensitivity to the proton radius is enhanced in kinematical regions associated with larger photon virtualities, in particular, at larger dilepton invariant mass, more forward or backward rapidities and larger transverse momenta of the outgoing protons. Theoretical predictions are further confronted with available ATLAS and CMS measurements of exclusive dilepton production in proton-proton collisions at $\sqrt{s}=7$ and $13$~TeV \cite{ATLAS:2015wnx,CMS:2011vma,ATLAS:2017sfe}. We assess the present experimental sensitivity to the proton radius parameter.

Within the LHC physics program, particle production in two-photon interactions in hadronic collisions has attracted increasing interest, both as a precision test of the Standard Model \cite{ATLAS:2024wla} and as a potential window to physics beyond the Standard Model \cite{Ogrodnik:2024bou,Thomas:2016ybw}.

\section{Proton Radius Puzzle and electromagnetic form factors}
\label{sec:proton}


The \emph{proton radius puzzle}  was historically associated with the difference between the proton charge radius extracted from elastic electron-proton scattering and ordinary-hydrogen spectroscopy, and the smaller value obtained from muonic hydrogen \cite{RevModPhys.94.015002,Carlson:2015jba,pohl2010size,Antognini:2013txn}. This interpretation is now more nuanced. Recent electron-scattering measurements and dispersive analyses of the proton electromagnetic form factors also support a smaller proton charge radius \cite{Xiong:2019umf,Lin:2021umz,Lin:2021xrc}. In the present work, the two radii are therefore used as benchmark values to quantify the sensitivity of exclusive photon-induced dilepton production to the proton electromagnetic form factor.

Our aim is not to provide a global reassessment of all available radius determinations. Instead, we adopt two representative proton charge radii frequently quoted in the context of the proton-radius puzzle, following the Particle Data Group (PDG 2018)~\cite{tanabashi2018review}. 
The determination using electron results in a \textit{larger} proton size, described by radius $r_L = 0.8751 \pm 0.0061~\mathrm{fm}$~\cite{Mohr:2015ccw}.
The determination based on muonic hydrogen gives a \textit{smaller} proton, with radius $r_S = 0.84087 \pm 0.00039~\mathrm{fm}$~\cite{Antognini:2013txn}.

The proton electromagnetic structure is not purely electric. It is described by the Sachs electric and magnetic form factors, $G_E(Q^2)$ and $G_M(Q^2)$, which constitute the standard framework for the analysis of elastic electron-proton scattering \cite{Sachs:1962zzc,Perdrisat:2006hj}. In the present context, the relevant form factor enters through the combination
\begin{equation}
    F(Q^2)=\frac{4m_p^2G_E^2(Q^2)+Q^2G_M^2(Q^2)}{4m_p^2+Q^2}
\end{equation}
which shows explicitly that the magnetic contribution becomes increasingly relevant away from the very low-$Q^2$ region. High-precision analyses of the proton form factors establish that the low-$Q^2$ region is particularly important for radius-sensitive studies \cite{RevModPhys.94.015002,bernauer2010high,Perdrisat:2006hj,A1:2013fsc}. Our final results are obtained using the full electromagnetic structure of the proton, even though the main physics focus remains the sensitivity to the proton-radius parameter. 

In phenomenological applications, it is convenient to introduce a dipole parametrization of the electric form factor and to use it as a controlled one-parameter framework for studying the sensitivity to the proton-radius input. We therefore adopt the standard dipole form 
\begin{equation}
    G_E(Q^2) = \frac{1}{\left( 1+Q^2/ \Lambda^2 \right)^2} \;,
    \qquad 
G_M(Q^2) = \mu_p G_E (Q^2) \;.
    \label{Eq.ff_dipole}
\end{equation}
Other descriptions of the proton electromagnetic structure are available, including dispersive analyses, global form factor fits, GPD-based parametrizations, and extractions from exclusive lepton-production observables \cite{Perdrisat:2006hj,Lin:2021umz,Lin:2021xrc,Goharipour:2024mbk,Goharipour:2025yxm,Moradi:2025pkp}. 

In addition, polarization-transfer measurements have shown that the scaled ratio $\mu_p G_E(Q^2)/G_M(Q^2)$ decreases with increasing $Q^2$, demonstrating that the electric and magnetic form factors do not have the same momentum dependence \cite{JeffersonLabHallA:1999epl}. For this reason, the commonly used dipole relation in Eq.~\ref{Eq.ff_dipole} is adopted here only as a baseline one-parameter framework.
It should not be understood as a precise global parametrization of the proton electromagnetic form factors.

In the dipole formula, the cutoff parameter $\Lambda^2$ is related to the proton charge radius through the standard slope relation \cite{Halzen1984QuarksAL}
\begin{equation}
    r_p^2 = -6 \left.\frac{\text{d} G_E (Q^2)}{dQ^2}\right|_{Q^2=0}
 \;,
\end{equation}
which, for the dipole parametrization, gives
\begin{equation}
    r_p^{2} =  \frac{12}{\Lambda^2} (\hbar c)^2 
\end{equation}
In the present work, $\Lambda^2$ is therefore treated as the parameter controlling the proton-radius input within the adopted dipole ansatz. The corresponding scenarios used in the calculations are listed in Table~\ref{tab:scenarios}, while the full electromagnetic structure of the proton is retained in the subsequent evaluation of the photon flux.
In addition to the small and large proton scenario, we also consider the \textit{conventional} dipole cut-off value of $\Lambda^2_C = 0.71~\text{GeV}^2$, which corresponds to the radius  $r_C \simeq 0.81~\mathrm{fm}$.
The radius and $\Lambda^2$ used in the following are summarised in Table~\ref{tab:scenarios}.

\begin{table}[!ht]
    \caption{Summary of the scenarios for proton radius and $\Lambda^2$ parameters.}
    \label{tab:scenarios}
    \centering
    \begin{tabular}{l c c c c c}
    \toprule
        \bfseries Scenario &
        \bfseries \textit{L} -- large proton  &
        \bfseries \textit{S} -- small proton &
        \bfseries \textit{C} -- conventional\\
        \midrule
        Proton radius & 
        $r_L = 0.8751$ fm &
        $r_S = 0.84087$ fm & 
        $r_C = 0.81$ fm \\
        Dipole cut off & 
        \quad $\Lambda^2_L = 0.6081\ \text{GeV}^2$ \quad &
        \quad $\Lambda^2_S = 0.6587\ \text{GeV}^2$ \quad &
        \quad $\Lambda^2_C = 0.71\ \text{GeV}^2$   \quad \\
         \bottomrule
    \end{tabular}
\end{table}

\begin{figure}[!ht]
    \begin{minipage}[!h]{0.49\textwidth}
\includegraphics[width=\textwidth]{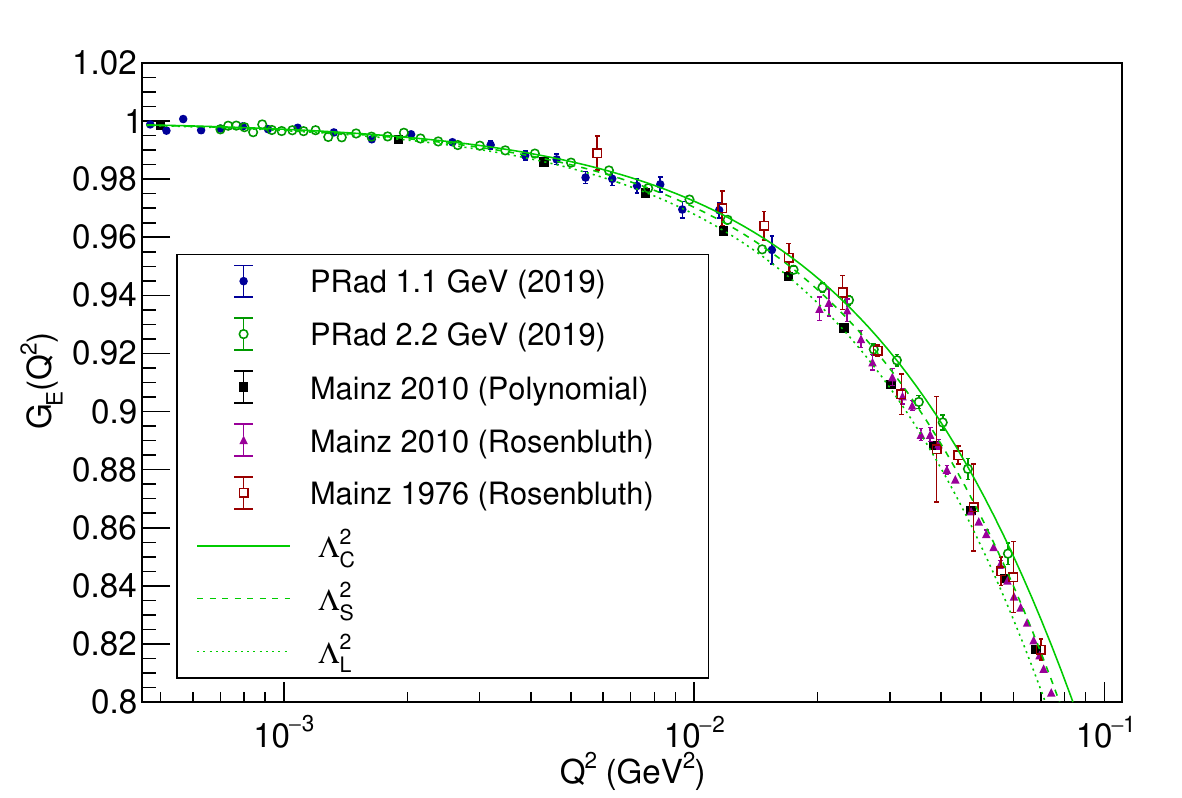}
    \end{minipage}
    \hfill
    \begin{minipage}[!h]{0.49\textwidth}
\includegraphics[width=\textwidth]{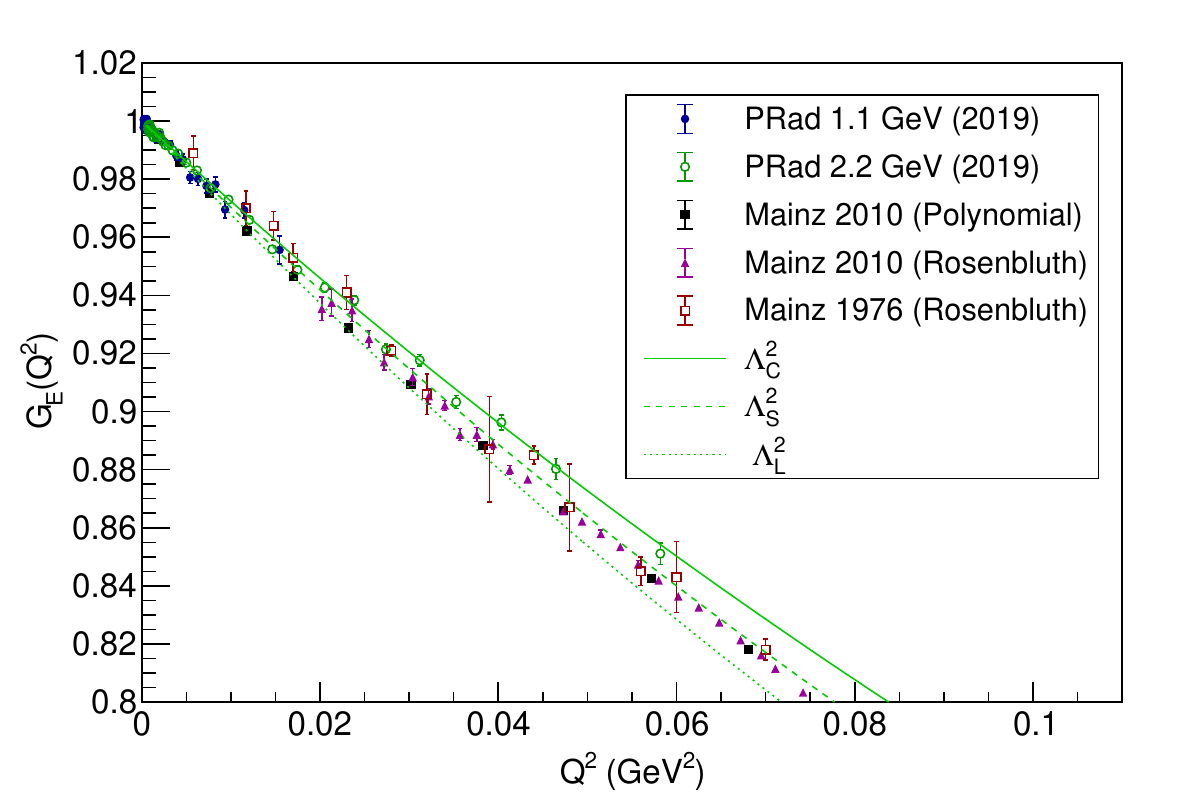}
    \end{minipage}
    \caption{Proton electric form factor as a function of $Q^2$, compared with PRad ($1.1$ and $2.2$~GeV) \cite{Xiong:2019umf}, and Mainz data \cite{Bernauer:2010zga,Hohler:1976ax}. The curves represent dipole parametrization for the $\Lambda^2$ values used in this study. These  $\Lambda^2$ scenarios cover the low-$Q^2$ region relevant to the proton-radius discussion. The left panel uses a logarithmic $Q^2$ scale, while the right panel presents the same data on a linear scale.}
\label{fig:formfactor_expt}
\end{figure}

Fig.~\ref{fig:formfactor_expt} compares the dipole form factors for three $\Lambda^2$ scenarios considered in this work to selected electron-scattering data.
The figure is not intended as a dedicated fit to form factor data, but rather as an illustration that the $\Lambda^2$ values considered here span a physically meaningful range for the subsequent analysis of photon-induced dilepton production.


Recent studies based on GPDs and exclusive lepton production also show that the proton charge and magnetic radii can be constrained through complementary electromagnetic observables \cite{Goharipour:2024mbk,Moradi:2025pkp}. These approaches are different from the two-photon mechanism studied here, but they provide useful context for using exclusive electromagnetic reactions as a process sensitive to the proton form factors.

\section{Theoretical framework for exclusive dilepton production}
\label{sec:theory}

\subsection{Momentum-space approach}
\label{sec:pspace}

\begin{figure}[!ht]
    \centering
    \includegraphics[scale=0.4]{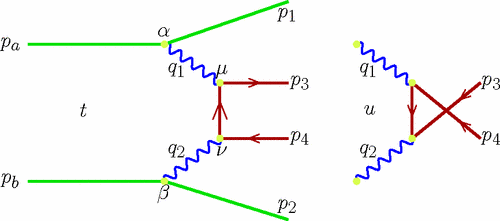}
    \caption{Feynman diagram for the exclusive dilepton production in proton-proton collision in $t$ (left panel) and $u$-channel (right panel). The momentum labels correspond to those used in Eqs.~\ref{eq:mom_space_1}-\ref{eq:mom_space_end}. Reproduced from Ref~\cite{Klusek-Gawenda:2010vqb})} 
    \label{fig:Feynman_2to4}
\end{figure}

Exclusive dilepton production in proton-proton collisions,
$pp \to pp\,\ell^+\ell^-$,
can be described in the momentum-space formalism as a $2\to 4$ process, with the lepton pair produced through a two-photon fusion. The corresponding production mechanism is illustrated in Fig. \ref{fig:Feynman_2to4}, which also defines the momentum assignment used throughout the paper: $p_a$ and $p_b$ denote the incoming proton momenta, $p_1$ and $p_2$ the protons in the final channel, and $p_3$ and $p_4$ the produced leptons. The proton electromagnetic structure is incorporated at the amplitude level via the proton--photon vertices.
The main advantage of the momentum-space approach is that it retains the full event kinematics and allows for a fully differential treatment of the final state. This is particularly useful when studying sensitivity to the proton radius across the phase space.

The total cross section for a $2\to4$ process can be evaluated in the momentum-space formalism as:
\begin{eqnarray}
\sigma(pp(\gamma\gamma)\rightarrow ppl^{+}l^{-}) & = & \int  \frac{1}{2s} \overline{ | \mathcal{M} | ^2} \left( 2 \pi \right)^4 \delta^4 \left( p_a+p_b-p_1-p_2-p_3-p_4 \right) \nonumber \\
      & \times & \frac{d^3 p_1}{\left( 2 \pi \right)^3 2E_1} \frac{d^3 p_2}{\left( 2 \pi \right)^3 2E_2} \frac{d^3 p_3}{\left( 2 \pi \right)^3 2E_3} \frac{d^3 p_4}{\left( 2 \pi \right)^3 2E_4} \; .
\label{eq:mom_space_1}
\end{eqnarray}
The momentum element in the cylindrical coordinates is given by $\frac{d^3p_i}{E_i}= dy_i p_{it}dp_{it} d\phi_i$, 
An auxiliary transverse momentum variable is defined as ${\bf p}_{mt} = {\bf p}_{3t} - {\bf p}_{4t}$. The expression, with the sum over lepton helicities included explicitly, is given by:
\begin{eqnarray}
\sigma(pp(\gamma\gamma)\rightarrow ppl^{+}l^{-}) & = & \int \sum_k \mathcal{J}_k^{-1} \left( p_{1t}, \, \phi_1, \, p_{2t}, \, \phi_2, \, y_3, \, y_4, \, p_{mt}, \, \phi_{p_{mt}} \right) \frac{1}{2\sqrt{s \left( s -4m^2 \right)}}   \nonumber \\
& \times & \overline{ | {\cal M}_{\lambda_3,\lambda_4} | ^2} \frac{1}{\left( 2 \pi \right)^8} \frac{1}{2^4} 
        \left( p_{1t} dp_{1t} d \phi_1 \right) 
      \left(p_{2t} dp_{2t} d \phi_2 \right) \frac{1}{4} dy_3 dy_4 d^2p_{m_t}  \; .
      \label{eq:mom_space_end}
\end{eqnarray}
where $\mathcal{J}_k$ is the Jacobian for the transformation from $\left( y_1, \, y_2 \right)$ to $\left( p_{1z}, \, p_{2z} \right)$. 
For the photon-induced processes, it is convenient to effect an additional change of variables, 
\textbf{$p_{it} \to \xi_i =\log_{10}\left(p_{it}\right)$}
which improves numerical convergence.

Contributions to the lepton-helicity amplitudes arise from the $t-$ and $u-$channel, as illustrated in Fig.~\ref{fig:Feynman_2to4}, and are expressed as:
\begin{equation}
{\cal M}_{\lambda_3,\lambda_4}={\cal M}_{\lambda_3,\lambda_4} \left( t \mbox{-channel} \right) + {\cal M}_{\lambda_3,\lambda_4} \left( u \mbox{-channel} \right),
\end{equation}
where
\begin{eqnarray}
 {\cal M}_{\lambda_3,\lambda_4} \left( t \mbox{-channel} \right) & = & e\, F\left( Q_1 \right) \, \left( p_a+p_1 \right)^\alpha
 \frac{-i\,g_{\alpha \mu}}{Q_1^2+i \varepsilon} \bar{u} \left( p_3,\, \lambda_3 \right) \, i \, \gamma^\mu
 \frac{i\, \left[ \left( \not{p}_3 - \not{Q}_1 \right) + m_{\ell} \right]}{\left( Q_1-p_3 \right)^2 -m_{\ell}^2} \nonumber \\
 & \times & i \, \gamma^\nu \,  v \left( p_4,\, \lambda_4 \right) \, \frac{-i\,g_{\nu \beta}}{Q_2^2+i \varepsilon} \, \left( p_b+p_2 \right)^\beta \, e\, F \left( Q_2 \right) 
 \label{eq:t_channel}
\end{eqnarray}
and
\begin{eqnarray}
 {\cal M}_{\lambda_3,\lambda_4} \left( u \mbox{-channel} \right) & = & e\, F \left( Q_1 \right) \, \left( p_a+p_1 \right)^\alpha
 \frac{-i\,g_{\alpha \mu}}{Q_1^2+i \varepsilon} \bar{u} \left( p_3,\, \lambda_3 \right) \, i \, \gamma^\nu
 \frac{i\, \left[ \left( \not{p}_3 - \not{Q}_2 \right) + m_{\ell} \right]}{\left( Q_2-p_3 \right)^2 -m_{\ell}^2} \nonumber \\
 & \times & i \, \gamma^\mu \,  v \left( p_4,\, \lambda_4 \right) \, \frac{-i\,g_{\nu \beta}}{Q_2^2+i \varepsilon} \, \left( p_b+p_2 \right)^\beta \, e\, F \left( Q_2 \right) \; .
 \label{eq:u_channel}
\end{eqnarray}
The amplitudes \ref{eq:t_channel} and \ref{eq:u_channel}, which depend on the form factors, are also determined numerically. The proton electromagnetic form factors enter explicitly via the factors $F(Q_1)$ and $F(Q_2)$ associated with the respective proton-photon vertices. Consequently, in the momentum-space calculation, the dependence on the proton-radius parameter is present at the amplitude level, rather than being introduced solely through an effective photon flux.
The integral in Eq.~\ref{eq:mom_space_end} is calculated numerically. 

In the present work, the momentum-space calculation is used as the reference formulation of the exclusive process. The corresponding amplitudes are evaluated numerically, and the multidimensional phase-space integration is performed explicitly. This exact treatment provides a benchmark against which the impact-parameter approach can be assessed.
This framework is especially important in the present context because it does not rely on the equivalent-photon approximation and therefore retains the full kinematic dependence of the intermediate-photon exchange. At the same time, absorptive effects are not naturally incorporated in this formulation, which motivates the complementary use of the impact-parameter representation discussed below \cite{Budnev:1975poe,Klusek-Gawenda:2010vqb}. 

\subsection{Impact-parameter space approach}

A complementary description of the exclusive process is provided by the impact-parameter representation, formulated within the Equivalent Photon Approximation (EPA) \cite{Budnev:1975poe,Williams:1934ad}. In the ultrarelativistic limit, the electromagnetic field of a fast-moving charged particle is dominated by its transverse component. As a result, the field of the incoming proton can be approximated as a flux of quasi-real photons, which constitutes the basis of the EPA used here. In this framework, the proton-proton reaction is expressed as a convolution of two photon fluxes with the elementary $\gamma \gamma \to \ell^+ \ell^-$ subprocess. When differential observables and fiducial cuts are considered, the calculation is performed using the differential elementary cross section as a function of the two-photon energy $W_{\gamma\gamma}$ and $z=\cos\theta$, where $\theta$ is the angle of the produced lepton with respect to the beam direction on the $\gamma\gamma$ center-of-mass frame.

\begin{figure}[!ht]
    \centering
    \begin{minipage}[b]{0.48\textwidth}
        \centering
        \includegraphics[width=6cm]{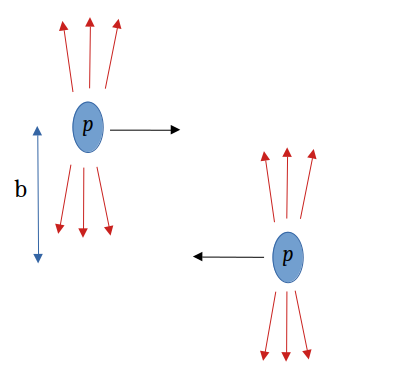}\\
        (a)
    \end{minipage}
    \hfill
    \begin{minipage}[b]{0.48\textwidth}
        \centering
        \includegraphics[width=6cm]{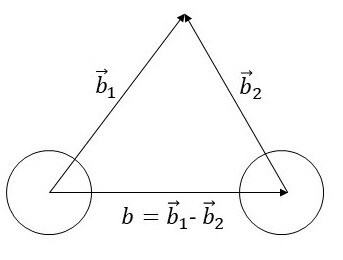}\\
        (b)
    \end{minipage}
    \caption{Schematic representation of the impact-parameter space formalism. Panel \textbf{(a)} shows the longitudinal projection of the proton--proton collision at impact parameter $b$. Panel \textbf{(b)} presents the transverse-plane projection, where the vectors $\vec{b_1}$ and $\vec{b_2}$ indicate the transverse distances from the centers of the respective protons to the photon--photon fusion (leptons creation) point.}
    \label{fig:impact_parameter}
\end{figure}

The geometric configuration is illustrated in Fig.~\ref{fig:impact_parameter}. The impact parameter of the proton--proton collision is denoted by the vector $\vec{b}$, while $\vec{b_1}$ and $\vec{b_2}$ represent the transverse distances from the centers of the protons to the photon--photon interaction point. This representation enables identification of dominant regions in transverse distance and of regions where rescattering effects are expected to be most important.

The full formula for the total $pp \to pp\, \ell^+ \ell^-$ cross section involves a six-dimensional integral: 
\begin{equation}
\begin{split}
    \sigma(pp \to pp\, \ell^+ \ell^-) &= 
    \int n(\omega_1, \vec{b}_1) \, n(\omega_2, \vec{b}_2) \,
    \frac{d\sigma_{\gamma\gamma \to \ell^+ \ell^-}}{dz}(W_{\gamma\gamma},z) \\ 
    &\quad \times \frac{W_{\gamma\gamma}}{2}\,
    {2\pi b\, db\, d\bar{b}_x\, d\bar{b}_y}\, dz dW_{\gamma\gamma}\, dY_{\ell^+\ell^-}
    \label{eq:sigma_tot_EPA}
\end{split}
\end{equation}
where $Y_{\ell^{+}\ell^{-}}= 1/2 (y_{\ell^{+}}+y_{\ell^{-}})$ is the rapidity of lepton pair. The factor $W_{\gamma\gamma}/2$ arises from the Jacobian of the transformation $(\omega_1,\omega_2)\rightarrow (W_{\gamma\gamma},Y)$, with $\omega_i = \frac{W_{\gamma\gamma}}{2} \exp (\pm Y_{\ell^{+}\ell^{-}})$. The transverse coordinates entering the photon fluxes are related to the proton--proton impact parameter through
\begin{equation}
    \vec{b}_1 = \left( \bar{b}_x + \frac{b}{2},\, \bar{b}_y \right)\;, \qquad 
    \vec{b}_2 = \left( \bar{b}_x - \frac{b}{2},\, \bar{b}_y \right) \;.
\label{eq:b1_b2_definition}
\end{equation}

The key object in this approach is the photon flux $n(\omega,b)$, which carries the full dependence on the electromagnetic structure of the emitting proton. In particular, the proton form factor enters directly in the flux, so that the sensitivity to the proton-radius input is already built into the impact-parameter formalism at this level, in contrast to the momentum-space approach, where this is done at the amplitude level.
The corresponding expression reads
\begin{equation}
    n(\omega,b) =  \frac{\alpha}{\pi^2 \omega} 
    \left[ \int dq_\perp \, q_\perp^2 
    \frac{G_{E}(Q^2)}{Q^2} \sqrt{\left(1 -\frac{\omega}{\gamma m_{p}}\right)\left(\frac{4m_p^{2}+Q^2\mu_p^{2}}{4m_p^{2}+Q^2}\right)}
    J_1(b q_\perp) \right]^2 \,.
    \label{eq:photonflux}
\end{equation}
Here, $Q^{2} = q_\perp^2 + \frac{\omega^2}{\gamma^2}$ and $\mu_p$ is the magnetic moment of the proton. 
Eq.~\ref{eq:photonflux} demonstrates that the magnetic contribution is retained in the flux, and its effect becomes significant at larger $Q^2$. 

\begin{figure}[!ht]
    \centering
    \includegraphics[scale=0.375]{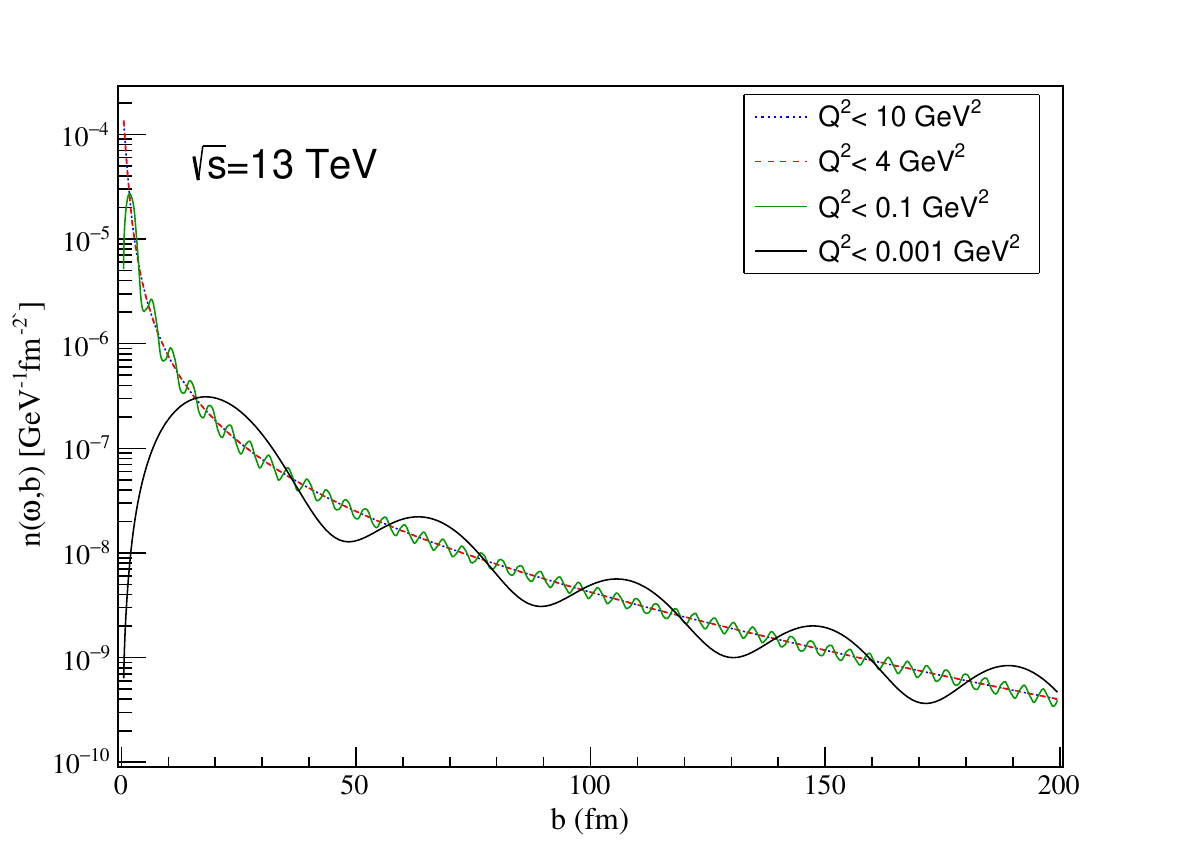}
    \caption{The photon flux $n(\omega,b)$ as a function of the impact parameter for various upper limits on the photon virtuality $Q^{2}$. The main differences appear at small $b$.}
    \label{fig:flux_Qsquare}
\end{figure}

This point is illustrated in Fig.~\ref{fig:flux_Qsquare}, where the photon flux is shown as a function of impact parameter for several upper limits of the photon virtuality. 
For very restrictive upper limits on $Q^2$, the flux exhibits visible oscillatory behavior as a function of $b$, which indicates that the virtuality range is still too limited to expose the full proton-structure effect. Once the upper limit is extended to $Q^2_{\max} \sim 4$–$10\,\text{GeV}^2$, these oscillations largely disappear and the physically relevant pattern becomes clear: the main differences are concentrated in the small-$b$ region. This observation supports the use of a sufficiently broad $Q^2$ range in the present analysis. 
In some studies \cite{Vysotskii:2018eic,Godunov:2023myj}, a low-$Q^2$ cutoff is considered ($Q^{2}<\Lambda^2_{QCD}$). In the present work, we retain the full electromagnetic structure of the proton in the flux, since this is required for a quantitatively reliable analysis of the sensitivity to the proton radius.

The impact-parameter representation is therefore used here as a framework in which the collision geometry and the role of rescattering corrections can be described in a natural way. Its validity is expected to be best when the exchanged photons are quasi-real, and their virtualities remain small compared with the invariant mass scale of the $\gamma\gamma \to \ell^+\ell^-$ subprocess. 

\subsection{Comparison of the two frameworks}

Before turning to the proton-radius dependence, it is useful to verify that the two theoretical frameworks lead to consistent predictions. To this end, we compare the momentum-space and impact-parameter calculations in a common kinematic setup: for the process $pp \to pp\,\tau^+\tau^-$ at $\sqrt{s}=13$~TeV, imposing the rapidity cut $|y_{\tau}|$< 3 and no additional restriction on the phase space.
\begin{figure}[!ht]
\includegraphics[width=0.49\textwidth]{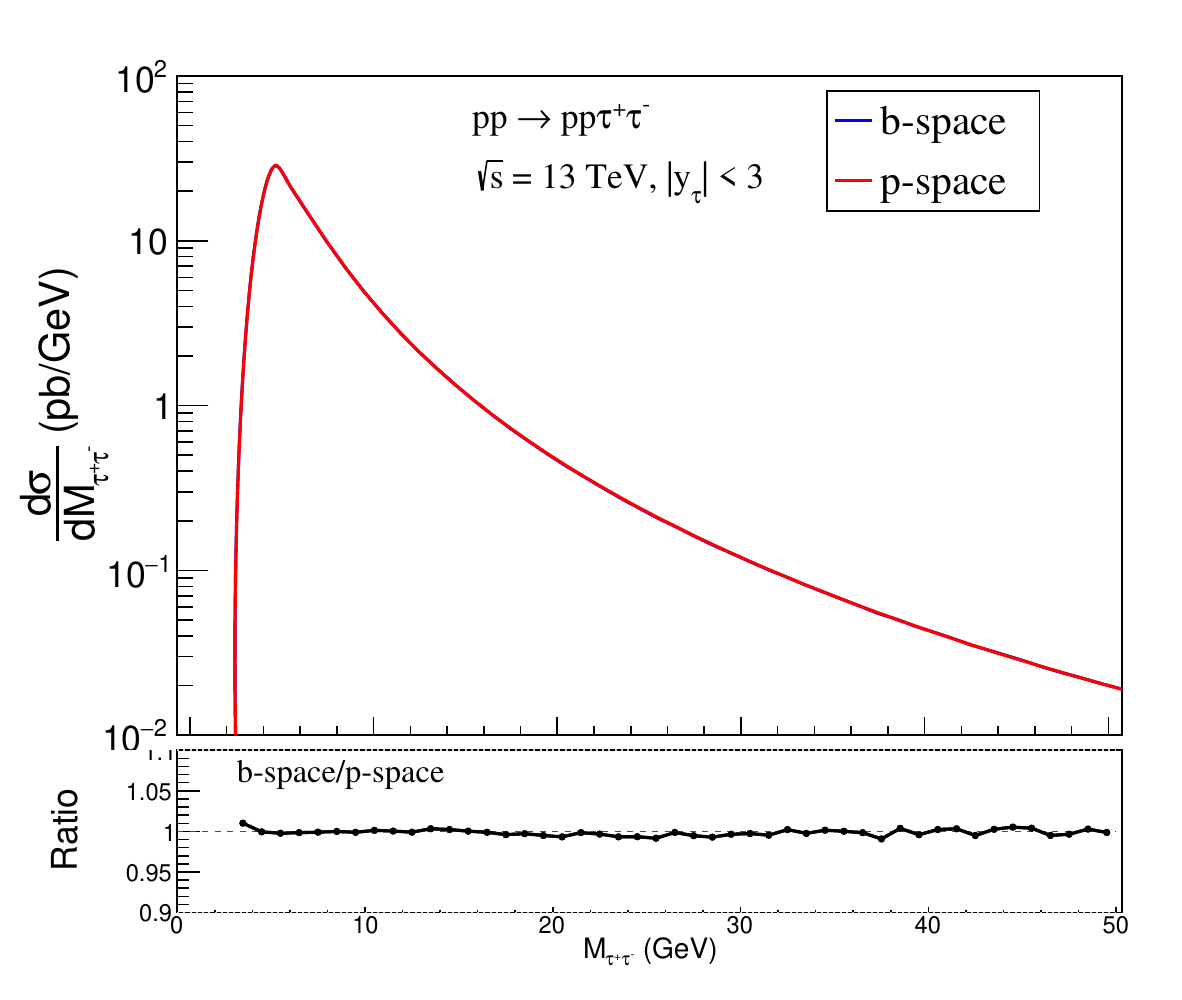}
    \caption{Differential cross section $d\sigma/dM_{\tau^+\tau^-}$ for the process $pp \to pp \tau^+\tau^-$ at $\sqrt{s}=13$~TeV, calculated in the momentum-space and impact-parameter formulations for the tauon rapidity range $|\eta_{\tau}|$< 3. The lower panel shows the ratio of the two results.}
    \label{fig:dsig_dM_b_and_b}
\end{figure}

The comparison is shown in Fig.~\ref{fig:dsig_dM_b_and_b} for the differential distribution $d\sigma/dM_{\tau\tau}$. The momentum and impact parameter space calculations are performed using Monte Carlo simulations based on the VEGAS adaptive algorithm \cite{Lepage:1977sw}. The momentum-space cross section yields a value of 112.1103 $\pm$ 0.002 pb. The corresponding impact-parameter result gives an integrated cross section of ${112.1435 \pm 0.007\,\text{pb}}$. The quoted uncertainty is the statistical precision of the MC evaluation. The relative difference between the two results is below 0.1\%. This level of agreement confirms that the impact-parameter formulation constitutes a reliable approximation in the kinematic regime relevant to the present study. At the differential level, the two approaches align well across the full invariant mass range shown in Fig.~\ref{fig:dsig_dM_b_and_b}. Only small deviations from unity are visible in the ratio plot, and they do not affect the subsequent phenomenological comparison.

\section{Sensitivity of the proton-radius input}
\label{sec:sensitifity_proton}

In this section, we quantify how the cross sections change when the conventionally used dipole parameter $\Lambda_C^2=0.71$~GeV$^2$ is replaced by the two benchmark values $\Lambda_S^2$ and $\Lambda_L^2$ corresponding to the representative proton-radius scenarios introduced in Sec.~\ref{sec:proton}. Our purpose is not to decide which of the currently quoted proton-radius values should be preferred, but rather to show explicitly how much the widely used conventional choice differs from the values motivated by present proton-radius determinations. The largest deviations are expected in kinematic regions that probe smaller impact parameters and larger photon virtualities. 

\begin{figure}[!h]
    \centering
\includegraphics[width=0.49\textwidth]{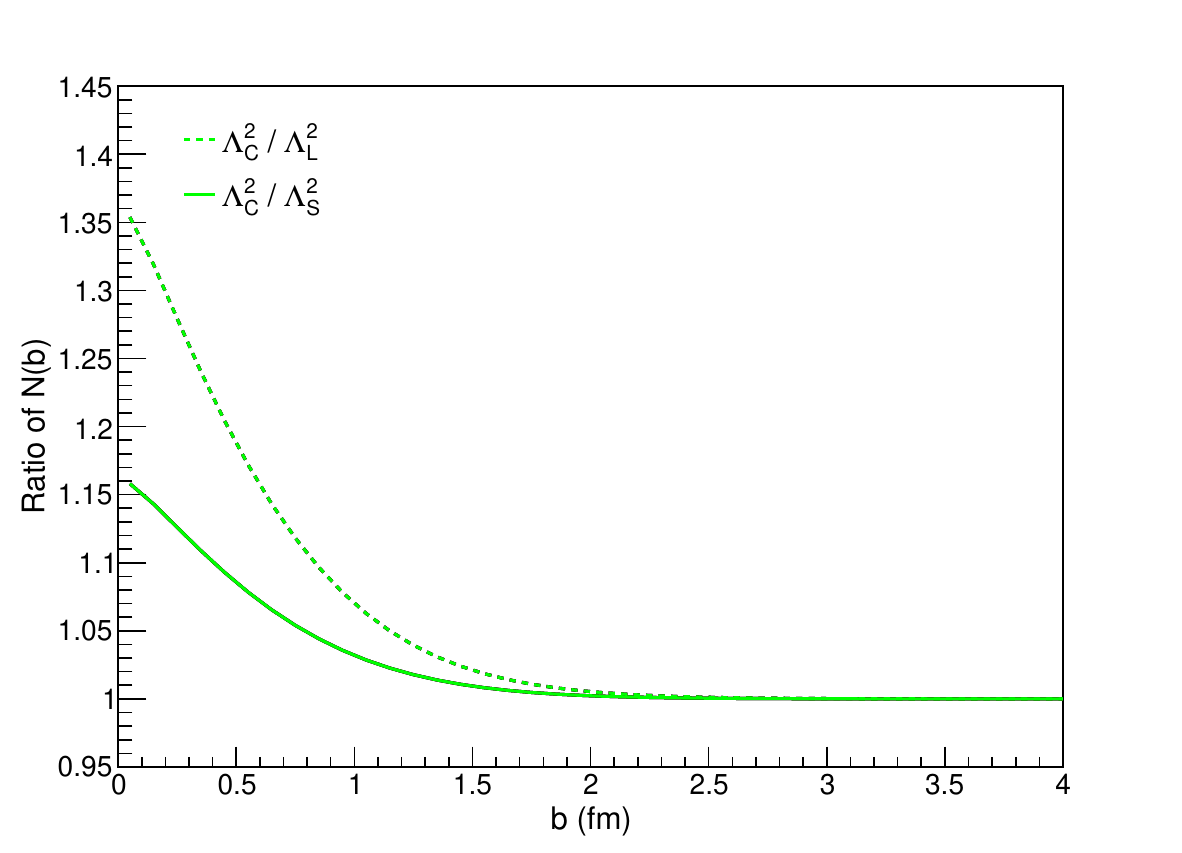}
    \caption{Sensitivity of the $pp \to pp \tau^+ \tau^-$ process at $\sqrt{s}=13$~TeV to the proton-radius choice. Ratio of the photon flux as a function of the impact parameter.}
    \label{fig:ratio_flux}
\end{figure}

Fig.~\ref{fig:ratio_flux} shows the ratio of the photon flux obtained with $\Lambda^2_C$ to that obtained with $\Lambda^2_L$ and $\Lambda^2_S$ as a function of the impact parameter $b$. The ratio decreases with increasing impact parameter and approaches a constant asymptotic value, indicating that at sufficiently large transverse distances, the photon flux no longer significantly resolves the proton size. The sensitivity is therefore concentrated in the small-$b$ region, where the proton's finite spatial structure is effectively probed. This observation also implies a stronger sensitivity to the assumed form factor parametrization. The same kinematic region that enhances the radius dependence probes larger photon virtualities. For this reason, the numerical size of the effect should be interpreted together with the form factor model-dependence study presented below.

This observation provides the basic physical interpretation for the differential results discussed below. Any observable that receives enhanced contributions from smaller impact parameters, or equivalently from larger photon virtualities, is expected to display a stronger dependence on the choice of $\Lambda^2$. Conversely, observables dominated by the large-$b$, low-$Q^2$ region should exhibit weaker sensitivity to the proton-radius input.

\begin{figure}[!h]
    \centering
    (a)\includegraphics[width=0.46\textwidth]{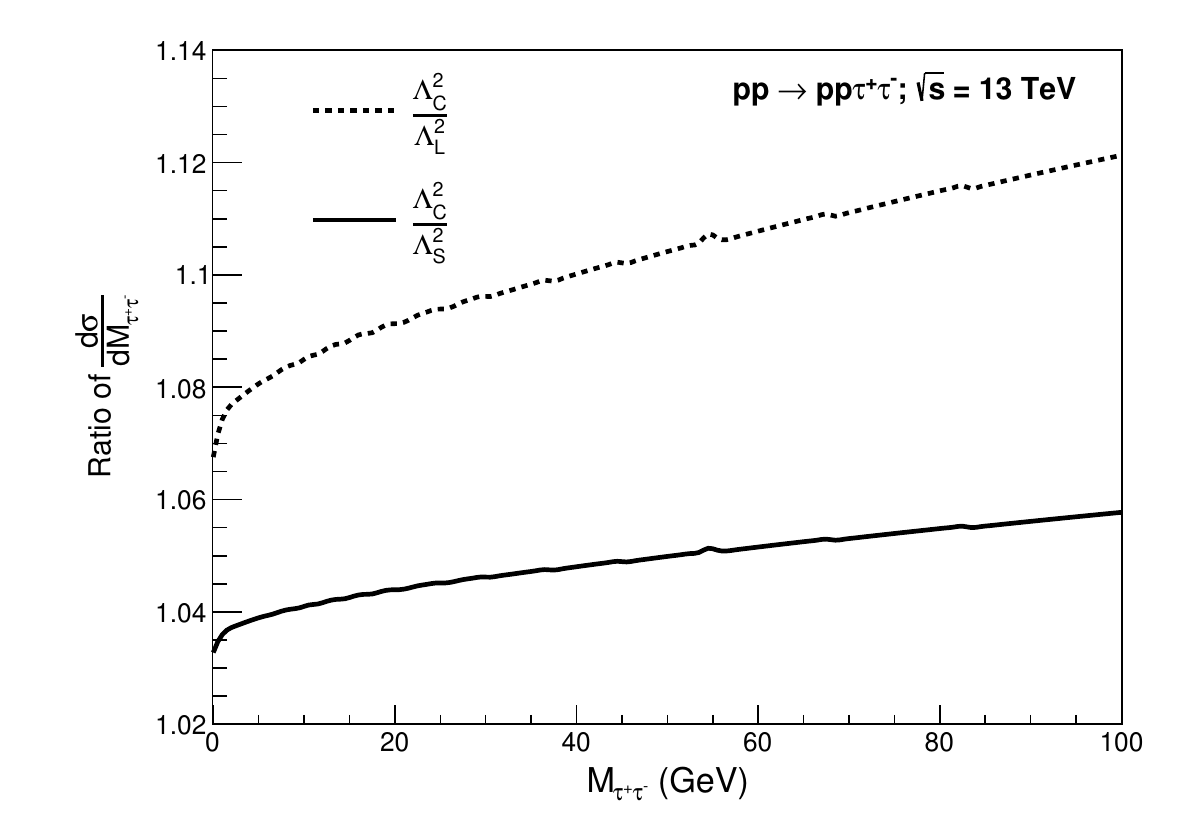}
    (b)\includegraphics[width=0.46\textwidth]{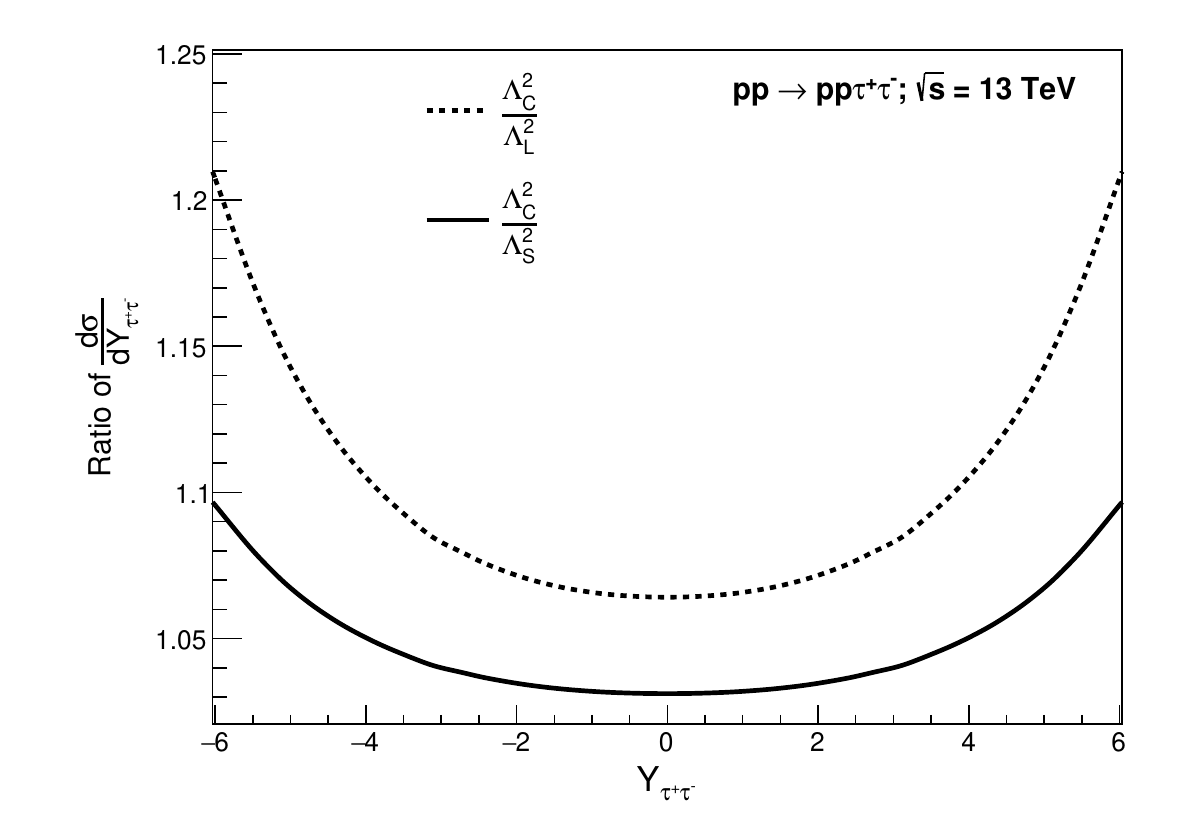}
    \caption{Ratios of the differential cross section for the $pp \to pp \tau^+ \tau^-$ process at $\sqrt{s}=13$~TeV, comparing the conventionally used dipole parameter $\Lambda_C^2$ with two benchmark values $\Lambda^2_S$ and $\Lambda^2_L$ corresponding to the proton-radius scenarios adopted in this work: (a) $d\sigma/dM_{\tau\tau}$ and (b) $d\sigma/dY_{\tau\tau}$. The curves correspond to the ratios $\sigma(\Lambda^2_C)/\sigma(\Lambda^2_L)$  (dashed line) and $\sigma(\Lambda^2_C)/\sigma(\Lambda^2_S)$  (solid line).
    }
    \label{fig:sigma_ratio}
\end{figure}

The effect is visible in Fig.~\ref{fig:sigma_ratio}, where we show the ratios of the differential cross sections obtained with $\Lambda_C^2$ to those obtained with $\Lambda_S^2$ and $\Lambda_L^2$, as functions of the dilepton invariant mass and ditauon rapidity in the process $pp \to pp\tau^+\tau^-$ at $\sqrt{s}=13$~TeV. In the invariant mass distribution, the ratio increases with $M_{\tau\tau}$, showing that the differences among the three $\Lambda^2$ choices become more sizeable at larger masses. In the rapidity distribution, the ratio is smallest around midrapidity and increases toward forward and backward rapidities, where the photon kinematics becomes more asymmetric and the sensitivity to the proton-radius input is enhanced. In both cases, the deviation is larger for the comparison with $\Lambda_L^2$, i.e., for the larger-radius benchmark, than for the comparison with $\Lambda_S^2$.

Indeed, the same trend is evident in the integrated cross sections. For the conventional choice $\Lambda_C^2=0.71$~GeV$^2$, we obtain $\sigma(\Lambda^2_C) = 140.47$~pb, compared with $\sigma(\Lambda^2_S) = 135.05$~pb for $\Lambda_S^2=0.6587$~GeV$^2$ and
$\sigma(\Lambda^2_L)=129.69$~pb for $\Lambda^2_L=0.6081$~GeV$^2$. Thus, relative to the PDG-motivated benchmarks, the conventional parameter yields larger total cross sections by about $4.0\%$ and $8.3\%$, respectively. This shows that the commonly used value $\Lambda^2=0.71$~GeV$^2$ is not a numerically neutral choice: even before absorptive corrections are included, it leads to systematic shifts in both the integrated and differential predictions.
The sensitivity discussed in this section should therefore be understood within the adopted dipole-based framework. In particular, the size of the effect may depend not only on the radius-setting parameter $\Lambda^2$, but also on how the magnetic contribution and absorption corrections, which become increasingly important in the same kinematic region, are accounted for.
These differences will be revisited in the next section after including rescattering effects.

\section{Rescattering corrections and their impact on proton-radius sensitivity}
\label{sec:corrections}

The sensitivity to the proton-radius input discussed in the previous section was obtained for the pure two-photon mechanism, i.e. for the QED subprocess $\gamma\gamma \to \ell^+ \ell^-$ set in proton--proton collisions. In realistic hadron interactions, however, additional soft hadronic interactions may occur between the colliding protons and destroy the exclusivity of the final state. The possibility of such extra interactions is accounted for through the survival factor $S^2_{\gamma\gamma}(b)$ interpreted as the probability that no additional soft $pp$ interaction takes place, Ref.~\cite{Dyndal:2014yea}. As a result, $S^2_{\gamma\gamma}(b)$ suppresses the contribution from the small impact parameter region, where protons overlap and hence the chance of extra inelastic activity is the largest. The corresponding parametrization,
\begin{equation}
    S^2_{\gamma\gamma}(b) = (1 - e^{-b^2/2B})^2  \;,
\end{equation}
is used together with the elastic-slope parameter $B$. Its value is taken from ATLAS measurements of elastic scattering at LHC energies \cite{ATLAS:2014vxr,ATLAS:2022mgx}. The cross section introduced by Eq.~\ref{eq:sigma_tot_EPA} completed with the absorption correction takes the form:

\begin{align}
    \sigma(pp \to pp\, \ell^+ \ell^-) &= 
    \int n(\omega_1, \vec{b}_1) \, n(\omega_2, \vec{b}_2) \,
     \frac{d\sigma_{\gamma\gamma \to \ell^+ \ell^-}}{dz}(W_{\gamma\gamma},z)  \\ \nonumber
    &\quad \times S^2_{\gamma\gamma}(b) \frac{W_{\gamma\gamma}}{2}\,
    {2\pi b\, db\, d\bar{b_{x}}\, d\bar{b_{y}}}\, dzdW_{\gamma\gamma}\, dY_{\ell^{+}\ell^{-}} \;.
    \label{eq:crosssec_bspace}
\end{align}
Since the proton-radius sensitivity is concentrated in the small-$b$ region, rescattering corrections are expected to reduce the size of the effect.

This expectation is confirmed by the integrated cross section for the process $pp \to pp\tau^+\tau^-$. After including absorptive corrections, we obtain $\sigma_{abs}(\Lambda^2_C)=129.2$~pb, $\sigma_{abs}(\Lambda^2_S)=125.06$~pb, $\sigma_{abs}(\Lambda^2_L)=120.8$~pb. Thus, relative to the two benchmark proton-radius scenarios, the conventionally used value $\Lambda^2_C=0.71$~GeV$^2$ still gives larger cross sections by about $3.3\%$ and $6.9\%$, respectively. For comparison, before absorptive corrections, the corresponding differences were about $4.0\%$ and $8.3\%$. Rescattering effects therefore, reduce the proton-radius sensitivity, as expected, but do not eliminate it.

\begin{figure}[!ht]
    \centering
    (a) \includegraphics[width=0.46\textwidth]{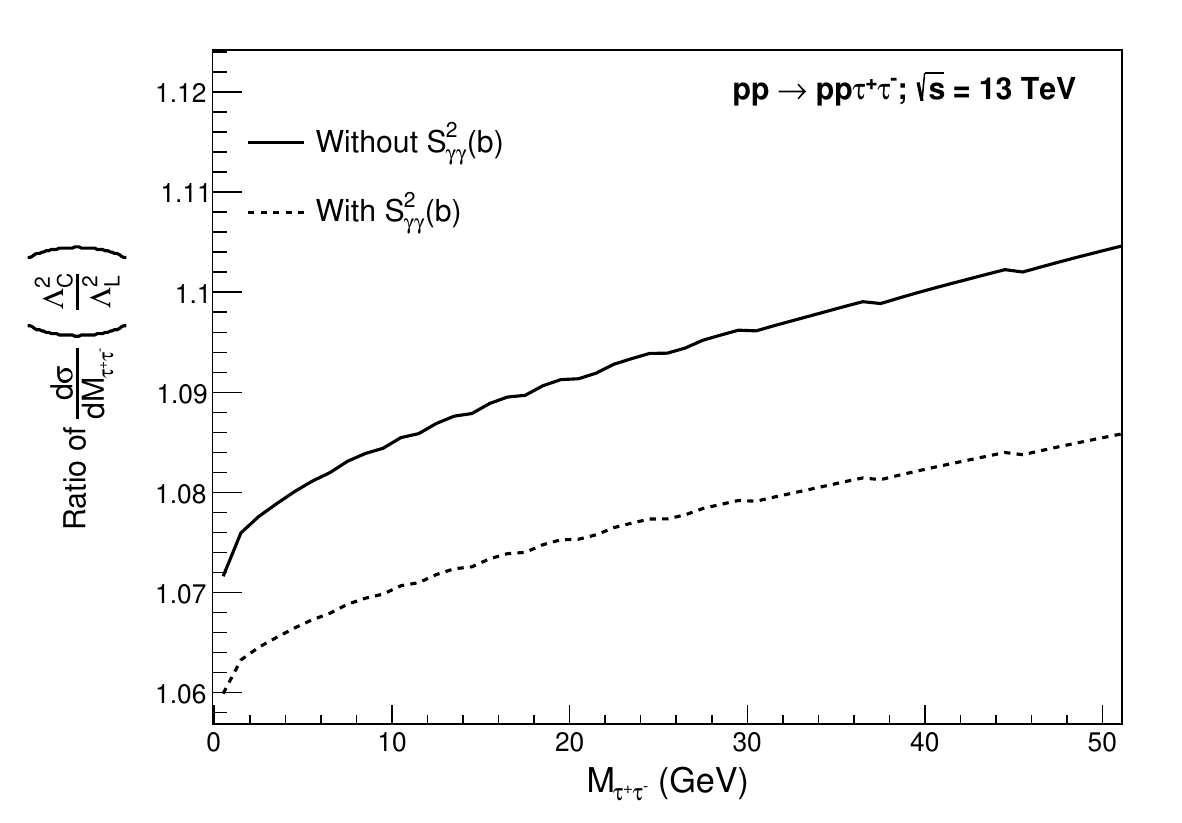}
    (b) \includegraphics[width=0.46\textwidth]{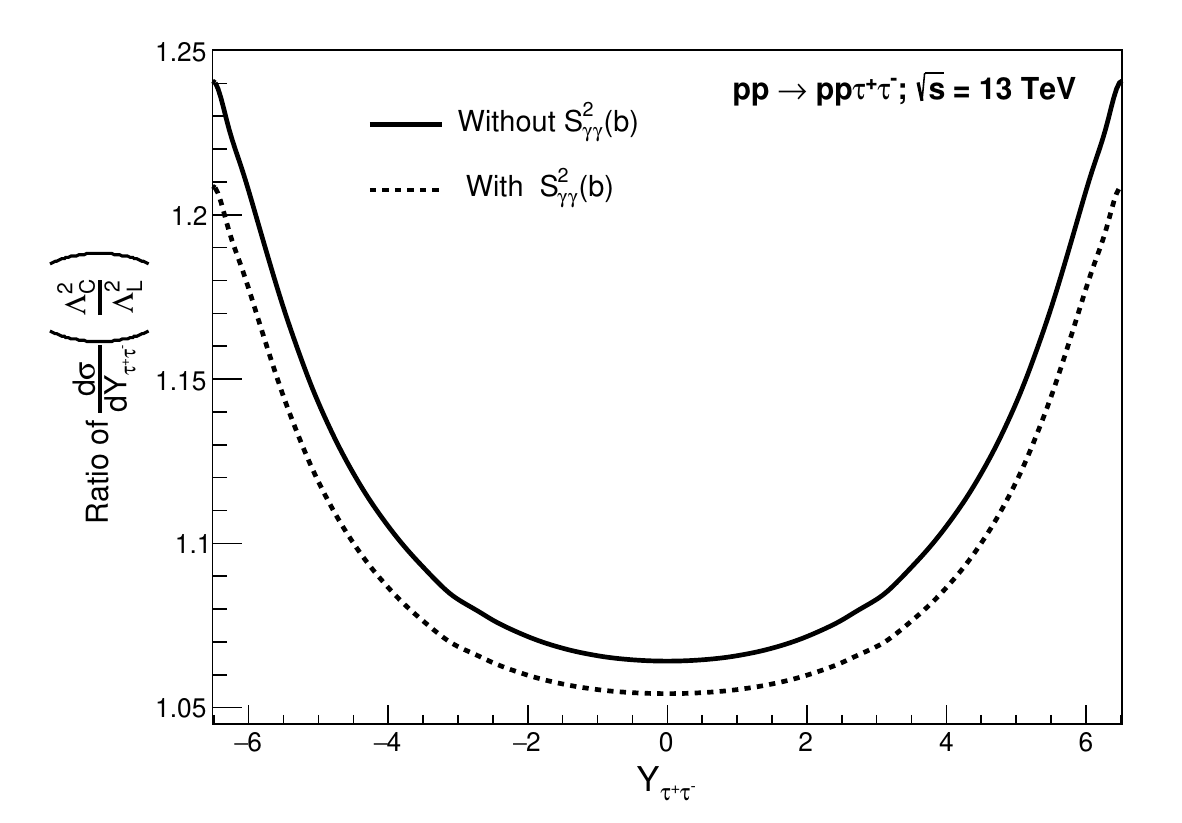}
     \caption{Ratios of the differential cross sections for the process $pp \to pp \tau^+\tau^-$ at $\sqrt{s}=13$~TeV, comparing results obtained with and without absorptive corrections: (a) $\frac{d\sigma}{dM_{\tau^{+}\tau^{-}}}$, (b) $\frac{d\sigma}{dY_{\tau^{+}\tau^{-}}}$.}
    \label{fig:ratio_diff_sigma_scattering}
\end{figure}

The same pattern is seen at the differential level. Fig.~\ref{fig:ratio_diff_sigma_scattering} shows the ratios obtained with and without the survival factor for the dilepton invariant mass and rapidity distributions. In both observables, the inclusion of absorptive corrections lowers the ratios. Nevertheless, the separation between the conventional dipole choice and the benchmark scenarios remains visible, showing that the radius dependence survives even after realistic rescattering effects are included.

Although the sensitivity analysis presented above is illustrated for the $\tau^+\tau^-$ channel, the underlying mechanism is general: an enhanced dependence on the proton-radius input emerges in kinematic regions probing larger photon virtualities and smaller impact parameters. For this reason, the subsequent phenomenological comparison is performed using the experimentally accessible $e^+e^-$ and $\mu^+\mu^-$ channels. We expect the same physical trend to persist in these cases, albeit within the specific fiducial regions defined by the ATLAS and CMS detector geometries. This approach allows us to confront the theoretical framework with existing LHC data while maintaining the generality of our conclusions regarding the proton electromagnetic properties.

\section{Comparison with LHC data and present experimental sensitivity}
\label{sec:results_LHC}

Next, we confront the theoretical predictions, including absorptive corrections, with the available LHC measurements of exclusive dilepton production. The purpose of this section is twofold. First, we verify that the present framework provides a realistic description of the measured fiducial cross section. 
Then, we study the sensitivity of the LHC data to the proton radius by performing a global fit of the theoretical model with the proton radius treated as a free parameter to the available LHC measurements.

The comparison of fiducial cross sections for the process $pp \rightarrow pp \ell^+\ell^-$  ($\ell=e^-,\mu^-$) with ATLAS and CMS measurements at $\sqrt{s}=7$ and $13$~TeV is summarized in Table~\ref{tab:cross_sections}.
All the presented theoretical values include the magnetic contribution in the full Sachs approach.
Taking only the electric form factor would reduce the fiducial cross sections by $6-9\%$, depending on the process and region. For example, in the ATLAS $13$~TeV muon channel, the full Sachs predictions, $\sigma(\Lambda^2_C)=3.622$~pb and $\sigma(\Lambda^2_L)=3.484$~pb, lie further above the measured value $3.12 \pm 0.07 \pm 0.14$~pb than the corresponding electric-only results, $\sigma(\Lambda^2_C)=3.373$~pb and $\sigma(\Lambda^2_L)=3.266$~pb. The difference is non-negligible in the context of precision comparisons to LHC data. For this reason, only the full Sachs structure results are used in the final phenomenological discussion. 
The fact that the electric-only approximation happens to lie closer to the central experimental values should not be interpreted as evidence that the magnetic contribution ought to be neglected. It rather shows that present fiducial data are not sufficient, by themselves, to isolate the proton-radius dependence from other normalization effects in the calculation.

\FloatBarrier
\begin{table*}[!h]
    \caption{Comparison of the theoretical fiducial cross section, calculated in the impact parameter framework with absorptive corrections and the full Sachs form factor treatment (for $\Lambda^{2}_{\text{C}}$ and $\Lambda^{2}_{\text{L}}$ separately), with ATLAS and CMS measurements at $\sqrt{s}=7$~TeV~\cite{ATLAS:2015wnx,CMS:2011vma} and $\sqrt{s}=13$~TeV~\cite{ATLAS:2017sfe}.
    The relative numerical uncertainties on the theoretical values are about $5\cdot10^{-5}$.}
    \label{tab:cross_sections}
    \centering
    \renewcommand{\arraystretch}{1} 
    \setlength{\tabcolsep}{0.3em}
    \vspace{1ex}
    \begin{tabular*}{\textwidth}{@{\extracolsep{\fill}} l c c c @{}}
    \toprule
        \textbf{Process \& Kinematics} & $\sigma_{\text{th}}(\Lambda^{2}_{\text{C}})$ (pb) & $\sigma_{\text{th}}(\Lambda^{2}_{\text{L}})$ (pb) & $\sigma_{\text{exp}}$ (pb) \\  
        \midrule

        \multicolumn{4}{l}{$pp \to pp \mu^+\mu^-$, $\sqrt{s} = 7$ TeV, ATLAS} \\
        \footnotesize $M_{\mu^+\mu^-}>20\;\text{GeV},\; p_T>10\;\text{GeV}$
        & 0.745
        & 0.708
        & $0.628 \pm 0.032 \pm 0.021$~\cite{ATLAS:2015wnx} \\[2ex]

        \multicolumn{4}{l}{$pp \to pp e^+e^-$, $\sqrt{s} = 7$ TeV, ATLAS}  \\
        \footnotesize $M_{e^+e^-}>24\;\text{GeV},\; p_T>12\;\text{GeV}$
        & 0.461
        & 0.438
        & $0.428 \pm 0.035 \pm 0.018$~\cite{ATLAS:2015wnx} \\[2ex]

        \multicolumn{4}{l}{$pp \to pp \mu^+\mu^-$, $\sqrt{s} = 7$ TeV, CMS}  \\
        \footnotesize $M_{\mu^+\mu^-}>11.5\;\text{GeV},\; p_T>4\;\text{GeV}$
        & 3.914
        & 3.753
        & $3.38^{+0.58}_{-0.55} \pm 0.21$~\cite{CMS:2011vma} \\[2ex]
        
        \multicolumn{4}{l}{$pp \to pp \mu^+\mu^-$, $\sqrt{s} = 13$ TeV, ATLAS}  \\
        \footnotesize $12<M_{\mu^+\mu^-}<70\;\text{GeV},\; p_T>6\;\text{GeV}$
        & 3.622
        & 3.484
        & 3.12 $\pm$ 0.07 $\pm$ 0.14~\cite{ATLAS:2017sfe}\\
   \bottomrule 
    \end{tabular*}
\end{table*}



\begin{figure}[!ht]
    \centering
\includegraphics[width=0.49\textwidth]{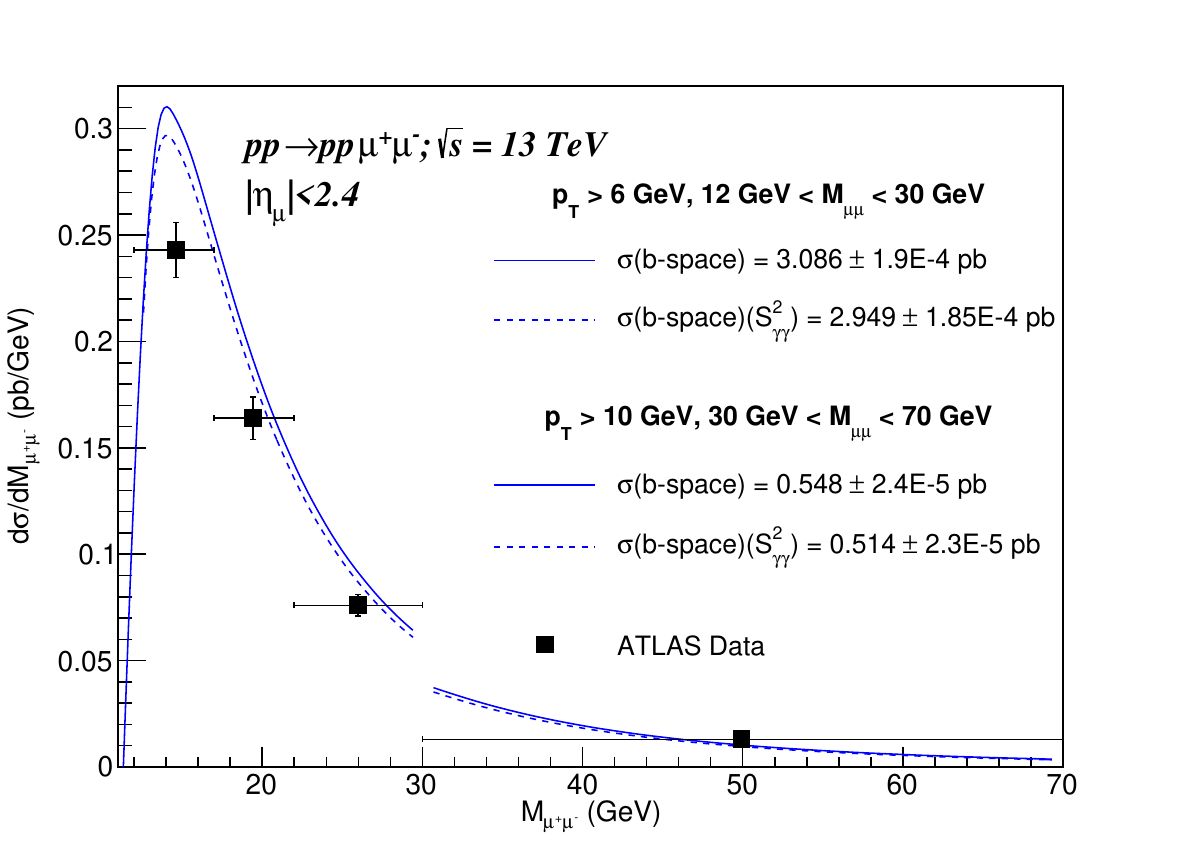}
    \caption{Differential cross section $d\sigma/dM_{\mu^+\mu^-}$, computed in b-space including survival factor. Results are calculated for ATLAS datasets, including kinematical cuts on $p_t$, $M_{\mu^+\mu^-}$ and $|y_\mu| < 2.4$. The dashed curves indicate the predictions after including the survival factor.}
    \label{fig:p_b_abseff}
\end{figure}

The situation becomes clearer at the differential level.
Fig.~\ref{fig:p_b_abseff} shows the $M_{\mu^+\mu^-}$ distribution in the ATLAS fiducial phase space at $\sqrt{s}=13$~TeV. In contrast to Table~\ref{tab:cross_sections}, this figure allows assessment of the mass spectrum shape in addition to the overall normalization. The survival factor suppresses the cross section by about $3.5\%$ in the lower-mass interval and about $4.9\%$ in the higher-mass interval. It is consistent with the stronger impact of rescattering effects in kinematics corresponding to smaller impact parameters.


\begin{figure}[!ht]
    \centering
    \includegraphics[width=0.65\textwidth]{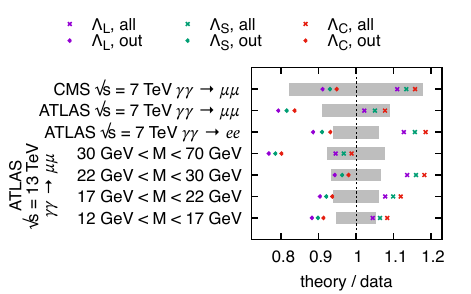}
    \caption{Ratios of theoretical results to ATLAS and CMS data. For the ATLAS $13$~TeV comparison, the theory curves have been averaged over the experimental bins of differential cross section $d\sigma/dM_{\mu^{+}\mu^{-}}$. The total cross section value is also compared in the case of CMS $7$~TeV data.
    }    
    \label{fig:rescattering_b}
\end{figure}

\par 
ATLAS collaboration has measured the above differential cross section for $\sqrt{s}= 13~\mathrm{TeV}$ data in each invariant mass bin from observed number of events normalized by integrated luminosity and bin width, resulting in a bin-averaged value~\cite{ATLAS:2017sfe}. In Fig.~\ref{fig:rescattering_b} we compare the ratio of our results to the ATLAS datapoints using three $\Lambda^2$ values over the same invariant mass bins as experimental measurements in the rapidity range $|\eta_{\mu}|<2.4$. We also compare our results with the fiducial cross section measured by ATLAS and CMS experiments for $\sqrt{s}= 7~\mathrm{TeV}$. 
The comparison is carried out by calculating the ratio of our calculations, including rescattering effects, for each experimental data point separately.
The experimental uncertainties are shown as the gray bands.
The calculations are performed for each of the three considered $\Lambda$ values ($\Lambda_S$, $\Lambda_L$ and $\Lambda_C$) and for two cases: \emph{all} -- calculations in the $b$-space including absorptive corrections as explained before, and $out$ -- production of photons inside protons is further suppressed by requiring $b_1 > r$ and $b_2 > r$, with $r$ corresponding to the $\Lambda^2$ value used in the particular calculation.

One can observe that none of the considered scenarios provides a good description of the data.
The \emph{all} calculations tend to overestimate the data, with the only exception being the highest-mass bin at 13~TeV. 
The \emph{out} scenario is always below the data.

To quantify the experimental sensitivity to the proton radius, we performed a $\chi^2$ fit of our theoretical predictions to the measurements of ATLAS and CMS.
The $\chi^2$ function was defined as:
\begin{equation}
    \chi^2(\Lambda^2) = \sum_i \frac{\left(\sigma_i^{\text{exp}}-\sigma_i^{\text{th}}(\Lambda^2)\right)^2}{\Delta_i^2},
\end{equation}
where
$i$ enumerates different data points,
$\sigma_i^\text{exp}$ are the measured cross section values,
and $\Delta_i$ is the combined experimental uncertainty.
\newcommand{\sigmath}[1]{\sigma_i^\text{th}(#1)}
The theoretical prediction for a given value of $\Lambda^2$, $\sigmath{\Lambda^2}$, is obtained by linear interpolation and extrapolation based on the complete calculations performed for $\Lambda^2 = \Lambda^2_C$ and $\Lambda^2 = \Lambda^2_L$:
\begin{equation}
    \sigmath{\Lambda^2} = \sigmath{\Lambda^2_C} + \left(\sigmath{\Lambda^2_L} - \sigmath{\Lambda^2_C}\right) \frac{\Lambda^2 - \Lambda^2_C}{\Lambda^2_L - \Lambda^2_C}.
\end{equation}

\begin{figure}[ht]
    \centering
    \includegraphics[width=0.65\textwidth]{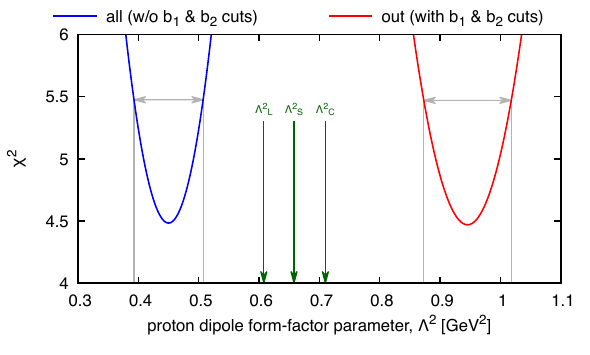}%
    \caption{The $\chi^2$ value as a function of $\Lambda^2$.}
    \label{fig:chi2}
\end{figure}

The shape of the $\chi^2(\Lambda^2)$ function, illustrating the compatibility between the data and theoretical predictions for different radii and both scenarios, \emph{all} and \emph{out}, is presented in Fig.~\ref{fig:chi2}.
Both scenarios give the best $\chi^2$ value of around 4.5, which for 6 degrees of freedom corresponds to $\chi^2/\text{ndf} = 0.75$ and p-value of about 0.60.
These values show a satisfactory level of agreement with the data.
However, the $\Lambda^2$ values of both minima are far away from the reasonable range.
This indicates some unknown problems in the theoretical description of the photon-induced processes and requires further studies.

On the other hand, one can appreciate the width of the $\chi^2$ parabolas, which turns out to be of a comparable magnitude to the spread between the three considered $\Lambda^2$ values. 
This shows that the LHC measurements of exclusive dilepton production are already sensitive to the proton charge radius at a level of a few percent.
The obtained result indicates a non-trivial sensitivity of exclusive dilepton data to the proton-radius scale
However, a precise determination of $r_p$ would require solving the apparent problem with the theoretical description.

At this point, it is useful to check how sensitive the results are to the assumption in Eq.~\ref{Eq.ff_dipole}, \textit{i.e.} that the electric and magnetic form factors have the same $Q^2$ dependence.
Following Ref.~\cite{Buchmann:2007zz} we use 
\begin{equation}
    R(Q^2) = \mu_p \frac{G_E^p(Q^2)}{G_M^p(Q^2)}=1 - 1.91 \frac{a \tau}{1+ d\tau} \, ,
\end{equation}
with $a=0.9$ and $d=2.8$.
The resulting fiducial cross sections are about 2--3\% larger than the nominal ones.
This leads to a conclusion that the photon-induced process $pp \to pp \ell^+ \ell^-$  is sensitive not only to the electric form factor, but also to the magnetic one.
However, the effects from changing $G_E^p$ and $G_M^p$ are very similar, and their separation is not possible, at least using the $pp \to pp \ell^+ \ell^-$  process considered here.
Some external constraint is needed, for example, on the $G_E^p/G_M^p$ ratio.

It should be pointed out that the available LHC data were not based on the full available luminosity.
In fact, the only LHC Run 2 measurement \cite{ATLAS:2017sfe} was based only on 3.2~fb$^{-2}$ of data taken at $\sqrt{s} = 13$~TeV.
The full dataset collected by ATLAS at this energy is almost 50 times larger and twice more data has been collected at $\sqrt{s} = 13.6$~TeV, and this is again doubled by CMS.
Future measurements with larger datasets, which could allow measurement for much higher invariant masses, or extended kinematic coverage, in particular -- the pseudorapidity range, could significantly improve the sensitivity. 
In addition, exploiting forward proton detectors to measure the proton transverse-momentum distributions could substantially improve the achievable precision of radius measurement.

\section{Conclusions}

Our study demonstrates that exclusive dilepton production in proton--proton collisions exhibits a sensitivity to the proton radius input within the dipole form factor framework. This sensitivity is minimal in the low-$Q^2$, large impact parameter regime, but becomes more pronounced in kinematics probing larger photon virtualities -- specifically at high dilepton invariant masses and away from midrapidity. While absorptive corrections reduce the overall magnitude of the effect, they do not eliminate it.

A central finding of our analysis is that the conventional dipole parameter ($\Lambda^2 = 0.71$~GeV$^2$) is not a numerically neutral reference choice. Replacing it with values corresponding to current proton-radius benchmarks leads to sizable shifts in both integrated and differential cross sections. Specifically, the conventional value consistently yields larger cross sections than the benchmark scenarios within the modern proton-radius range. This discrepancy persists even after accounting for absorptive corrections.

Crucially, we find that a more formally complete treatment of the proton electromagnetic structure does not necessarily improve agreement with available fiducial data. Including the magnetic contribution via the full Sachs form factors increases the predicted cross sections, moving them further from the ATLAS and CMS central values, whereas the electric-only approximation remains numerically closer to the data.

A fit to the available LHC measurements yields
\begin{equation}
r_p = 1.002 \pm 0.038 \text{ fm}, \qquad \Lambda^2 = 0.465 \pm 0.056  \text{ GeV}^2,
\end{equation}
with $\chi^2_{\min}=4.60$ and $\chi^2/\text{ndf}=0.77$. The central value aligns with the larger-radius side of the currently debated range. However, this result should not be over-interpreted as a precision extraction of the proton radius. Instead, it serves as an indicator of the radius scale preferred by the present implementation of the framework when confronted with current LHC data.

As a cross-check of the interpolation procedure, we also performed the calculation directly at the value preferred by the fit ($\Lambda^2 = 0.465~\mathrm{GeV^2}$). The resulting fiducial cross sections are very close to the measured values. For the ATLAS ($13$~TeV) muon measurement we obtain ($\sigma_{\mathrm{th}} = 3.259 \pm 8.8E-4~\mathrm{pb}$), compared with ($\sigma_{\mathrm{exp}} = 3.12 \pm 0.07 \pm 0.14~\mathrm{pb}$). For the ATLAS and CMS ($7$~TeV) muon measurements, the corresponding theoretical values are ($0.648 \pm 1.2E-4$~pb) and ($3.489 \pm 6.5E-4$~pb), to be compared with the measured values ($0.628 \pm 0.032 \pm 0.021$~pb) and ($3.38^{+0.58}_{-0.55} \pm 0.21$~pb), respectively.
The ($\chi^2$) value obtained from the direct calculation at ($\Lambda^2 = 0.465~\mathrm{GeV}^2$) is close to, but not identical with, the value obtained from the interpolated dependence on ($\Lambda^2$). This small difference reflects the fact that the cross sections are not perfectly linear functions of the dipole parameter. Nevertheless, the directly calculated value gives a lower ($\chi^2$) than the benchmark scenarios based on the conventional and PDG-motivated values of ($\Lambda^2$). 
This indicates that the value ($\Lambda^2 = 0.465~\mathrm{GeV}^2$) provides a better overall description of the data than the standard benchmark choices

To transform this observable into a precision tool for measuring the proton charge radius, we first need a thorough check of our theory systematics. Specifically, we must see how stable the extracted radius remains when we change how the magnetic contribution is handled, how the survival factor is modeled, or which form factor parametrization we choose.
The result should not be over-interpreted as a direct measurement of the proton charge radius. Rather, it demonstrates that exclusive dilepton production at the LHC is already sensitive to radius-scale variations in the proton electromagnetic structure. Turning this sensitivity into a precision determination will require dedicated experimental analyses and a consistent treatment of the associated theory uncertainties.

The reach of this probe will be significantly extended at future facilities. While the LHC provides the first constraints, the vastly larger kinematic range of the FCC-hh \cite{FCC:2018vvp} will probe the dipole structure at unprecedented $Q^2$. Conversely, measurements at RHIC \cite{Xie:2023vfi} and the EIC \cite{Chwastowski:2022fzk} offer a complementary window into the low-$Q^2$ limit and the validity of the photon flux modeling. Future progress requires not only higher statistics, but a refined theoretical treatment of unitarization to transform exclusive dilepton production into a precision probe of proton structure.

\section*{Acknowledgements}
This work was partially supported by the Polish National Science Center under grant No. 2021/42/E/ST2/00350.
\bibliographystyle{unsrturl}
\bibliography{bib}

\end{document}